\documentstyle[11pt]{article}
\begin{document}
\centerline{\Large\bf Teleportation of the Relativistic Quantum Field}
\vskip 3mm
\centerline{R.Laiho$^{\dag}$, S.N.Molotkov$^{\ddag}$, S.S.Nazin$^{\ddag}$}
\vskip 2mm
\centerline{\it\small $^{\dag}$Wihuri Physical Laboratory,}
\centerline{\it\small Department of Physics, University of Turku, 20014 Turku, Finland}
\centerline{\sl\small $^{\ddag}$Institute of Solid State Physics of
Russian Academy of Sciences,}
\centerline{\sl\small Chernogolovka, Moscow District, 142432 Russia}
\vskip 3mm

\begin{abstract}
The process of teleportation of a completely unknown one-particle
state of a free relativistic quantum field is considered.
In contrast to the
non-relativistic quantum mechanics, the teleportation of an unknown
state of the quantum field cannot be in principle described in terms
of a measurement in a tensor product of two Hilbert spaces to
which the unknown state and the state of the EPR-pair belong.
The reason is of the existence of a cyclic (vacuum) state common to both
the unknown state and the EPR-pair. Due to the common vacuum vector and the
microcausality principle (commutation relations for the field operators),
the teleportation amplitude contains inevitably contributions
which are irrelevant to the teleportation process. Hence in the
relativistic theory the teleportation in the sense it is understood in
the non-relativistic quantum mechanics proves to be impossible
because of the impossibility of the realization of the appropriate
measurement as a tensor product of the measurements related to the
individual subsystems so that one can only speak of the
amplitude of the propagation of the field as a whole.
\end{abstract}
\vskip 3mm

One of the fundamental results of the non-relativistic quantum
information theory consists in the possibility of the
teleportation of an unknown quantum state by means of a quantum
communication channel realized by a non-local entangled state
(an EPR-pair [1]) used together with a classical communication
channel [2].

When the unknown quantum state $|\psi\rangle_s$ belongs to a
finite-dimensional Hilbert space (dim${\cal H}_s<\infty$), the
teleportation can be preformed ideally and with the unit probability
employing an EPR-pair with finite energy. On the other hand,
if the state space of the teleported system is infinite-dimensional,
(dim${\cal H}_s=\infty$), the ideal teleportation formally requires
an EPR-pair with infinite energy [3,4]. However, the teleported
state can be made arbitrarily close to the input unknown state with the
probability arbitrarily close to unit by increasing the energy
of the EPR-pair and thus making the EPR correlations
more and more close the ideal ones.

The non-relativistic quantum mechanics yields only an approximate
description of the reality. A more correct and complete description
is provided by quantum field theory
(since the relativistic
quantum mechanics does not allow any sensible physical interpretation,
the relativistic theory arises from the very beginning as a quantum
field theory). Therefore, it is interesting to consider the
possibility of teleportation of a completely unknown state of the
relativistic quantum field. In addition, although all the teleportation
experiments carried out so far are performed with photons
which are essentially relativistic particles, they are always interpreted
within the framework of the non-relativistic quantum mechanics.

In the rest of the paper, considering a simple example, we shall
show that in the relativistic quantum field theory the teleportation of
even a one-particle state of free quantum field cannot be achieved
in the sense it is understood in the non-relativistic quantum mechanics.
The latter actually follows from the existence of a vacuum (cyclic)
state which is common to both the completely unknown state to be teleported
and the EPR pair together with the microcausality principle
(commutation or anticommutation relations for the field operators).

For convenience we shall briefly remind the teleportation procedure
in the non-relativistic case.

Suppose that we are given an unknown quantum state
$|\psi\rangle_s\in {\cal H}_s$ (dim${\cal H}_s<\infty$)
and a maximally entangled EPR-pair
$|\psi\rangle_{EPR}\in {\cal H}_{12}={\cal H}_{1}\otimes{\cal H}_{2}$
(dim${\cal H}_{12}<\infty$). The EPR-pair is a composite system
consisting of two particles with the state spaces ${\cal H}_{1}$ and
${\cal H}_{2}$. To achieve the teleportation, one performs a joint
measurement on the unknown quantum state and one of the particles
of the EPR-pair described by an identity resolution in
${\cal H}={\cal H}_{s}\otimes{\cal H}_{12}$ over some measurable outcome
space ${\Theta}$ (which is discrete, $\Theta=\sum_i\theta_i$, for
the teleportation of the states belonging to finite-dimensional spaces).
The measurement is defined by the identity resolution
\begin{equation}
I_{s12}=\sum_i {\cal M}(\theta_i)=\sum_i I_1 \otimes  {\cal M}_{2s}(\theta_i).
\end{equation}
If the measurement yields the $i$-th outcome, the subsystem 1 is
found in a new state
\begin{equation}
\rho^{i}_{1}=
\frac{\mbox{Tr}_{2s}\{ {\cal M}(\theta_i) ( \rho_{s} \otimes \rho_{EPR} ) \} }
{\mbox{Tr}_{12s}\{ {\cal M}(\theta_i) ( \rho_{s} \otimes \rho_{EPR} ) \} }.
\end{equation}
To within a unitary rotation $U_i$ which only depends
on the measurement outcome $i$ and does not depend on the unknown state,
the state (2) coincides with the unknown input state:
\begin{equation}
\tilde{\rho}_1 = \rho_s \quad \tilde{\rho}_1= U_i \rho^{i}_{1} U^{-1}_{i}  .
\end{equation}
The non-relativistic teleportation procedure substantially employs the fact
that the Hilbert state space of each subsystem can be accessed separately.

In the non-relativistic quantum mechanics physically different systems
are treated in the same way in the sense that any two systems are formally
considered to be identical if their state spaces are identical (isomorphic).
Therefore, formally, teleported is the unknown state vector rather than
the particle itself. There are no rules that prohibit the superposition
of the states belonging to physically different subsystems. In the quantum
field theory the situation is completely different.

Let us now turn to the teleportation of an unknown state of a free
quantum field. For simplicity we shall first consider the teleportation
of a one-particle state of the free scalar quantum field, although
the most interesting is perhaps the case of the gauge (photon) field.
To avoid unnecessary technical details associated with the indefinite
metrics, we shall restrict our analysis to the scalar field. All the
remarks concerning the scalar field teleportation are also relevant
to photon teleportation [9] (we mean the teleportation of a
completely unknown one-photon state when not only the polarization
state but also the wave packet shape is unknown).

The states of a relativistic quantum system are described by the
rays in the physical Hilbert space ${\cal H}$ where a
unitary representation of the covering Poincar\'e group is realized [5,6].
The local quantum field
$\varphi(\hat{x})$ (here $\hat{x}=(t,{\bf x})$ is a point in the
Minkowski space-time) is defined as a tensor (if the field has more
than one component) operator-valued distribution.  To be more precise,
corresponding to any function (or a set of functions, if the field is
multicomponent) $f(\hat{x})\in {\cal J}(\hat{x})$, where ${\cal
J}(\hat{x})$ is the space of test infinitely differentiable functions
decreasing together with all their derivatives at the infinity faster
than any polynomial [5,6]), is the operator symbolically written as
\begin{equation}
\varphi(f)=\sum_{j=1}^{r}\varphi_j(f_j)=
\sum_{j=1}^{r}  \int \varphi_j(\hat{x}) f_j(\hat{x}) d\hat{x}.
\end{equation}
The operators $\varphi(f)$ and $\varphi^*(f)$ have a common
domain which does not depend on $f(\hat{x})$, is dense in ${\cal H}$, and
is invariant under the action of the field operators,
$\varphi(f)\Omega\subset\Omega$ ($\varphi^*(f)\Omega\subset\Omega$).
For any vectors $|\phi\rangle,{} |\psi\rangle\in \Omega\subset {\cal H}$
the quantity $\langle\phi|\varphi(f)|\psi\rangle$ is a distribution
from ${\cal J}^*(\hat{x})$ (${\cal J}^*(\hat{x})$ is the space
of distributions conjugate to ${\cal J}(\hat{x})$).

The space $\Omega$ contains a cyclic vector, called the vacuum state,
$|0\rangle\in\Omega$ such that the set of all polynomials
${\cal P}(\varphi,f)$ constitute an operator algebra with involution whose
action on $|0\rangle\in\Omega$ generates the entire space $\Omega$.
The field operator algebra is defined as
\begin{equation}
{\cal P}(\varphi,f)=f_0+\sum_{n=1}^{\infty}\int\int\ldots\int
\varphi(\hat{x}_1)\varphi(\hat{x}_2)\ldots\varphi(\hat{x}_n)
f(\hat{x}_1,\hat{x}_2,\ldots, \hat{x}_n)
d\hat{x}_1 d\hat{x}_2 \ldots d\hat{x}_n,
\end{equation}
\begin{displaymath}
f(\hat{x}_1,\hat{x}_2,\ldots, \hat{x}_n) \in {\cal J}(\hat{x}^n).
\end{displaymath}

The fact that the field operators form an algebra implies that
any observable can be expressed through the field operators [5,6].

The unsmeared field operators $\varphi(\hat{x})$ map the regular states
from $\Omega$ to the generalized states
${\cal P}(\varphi(\hat{x}))\Omega\subset\Omega^*$
($\Omega^*$ is the conjugate space to $\Omega$ consisting of all the
linear functionals defined on $\Omega$ and continuous with respect to
the scalar product in ${\cal H}$).
The microcausality principle is also postulated; to be more precise,
the field operators are assumed to commute (anticommute) if the
supports of their corresponding functions $f(\hat{x}), g(\hat{y})$
are separated by a space-like interval
($\mbox{supp}f(\hat{x})\cdot g(\hat{y})\in(\hat{x}-\hat{y})^2<0$),
i.e. for any vector $|\psi\rangle\in\Omega$ we have
\begin{equation}
[\varphi(f),\varphi(g)]_{\pm}|\psi\rangle=0,\quad
(\hat{x}-\hat{y})^2<0.
\end{equation}
The expression (6) is interpreted as the impossibility of any
causal relation between the measurements performed in the domains
separated by a space-like interval since no interaction can propagate
faster than light.

Further, the requirements that the system states are described
by the rays in the Hilbert where a Poincar\'e group representation is
realized and the spectrum of the group generators in the momentum
representation lies in the front part of the light cone imply that
the Lorentz-covariant quantum field can only be realized as an
operator valued distribution rather than the field of operators
$\varphi(\hat{x})$ acting in ${\cal H}$ [5,6].

The interpretation of a quantum field as a field of operators acting in
${\cal H}$ is only consistent with the trivial two-point function
$\langle 0|\varphi^-(\hat{x})\varphi^+(\hat{y})|0\rangle=const$ and
results in an obvious violation of the microcausality principle.
The smearing function $f(\hat{x})$ can be interpreted (with some reservations)
as the amplitude (``shape'') of the one-particle packet.

Let us now construct the EPR-state, the one-particle state of the scalar
to be teleported, and the corresponding measurement for the relativistic case
emphasizing the differences from the non-relativistic theory.

The state of the EPR-pair is described by the vector
$|\psi\rangle_{EPR}\in\Omega\in{\cal H}$ in the subspace of the two-particle
states. The most general form of the relevant vectors is
\begin{equation}
|\psi\rangle_{EPR}=
{\cal P}_2(\varphi,{\cal F})|0\rangle = \int\int d\hat{x}_1 d\hat{x}_2
{\cal F}(\hat{x}_1,\hat{x}_2) \varphi^+(\hat{x}_1) \varphi^+(\hat{x}_2)
|0\rangle,
\end{equation}
where
\begin{displaymath}
\varphi^{\pm}(\hat{x})=\frac{1}{(2\pi)^{3/2}}
\int \mbox{e}^{\mp i\hat{k}\hat{x}} \theta(k^0)\delta(\hat{k}^2-m^2)
a^{\pm}(\hat{k}) d\hat{k}=
\frac{1}{(2\pi)^{3/2}} \int_{V^+_m}
\mbox{e}^{\mp i\hat{k}\hat{x}} a^{\pm}({\bf k})
\frac{d{\bf k}}{\sqrt{2k^0} },
\end{displaymath}
\begin{displaymath}
\hat{k}\hat{x}=k^0x^0-{\bf k}\cdot{\bf x},\quad
k^0=\sqrt{{\bf k}^2 + m^2 }.
\end{displaymath}
The symbol $V_m^+$ in the second integral is kept to emphasize the fact
that contributing to the integral are only the values at the mass shell
inside the front part of the light cone $k^0>0$.
The ideal EPR correlations correspond to the case where
\begin{equation}
\tilde{{\cal F}}(\hat{x}_1,\hat{x}_2)=\delta(x_1^0-x^0) \delta(x_2^0-x^0)
\delta({\bf x}_1-{\bf x}_2) const({\bf x}_1+{\bf x}_2).
\end{equation}
However, the function (8) does not belong to the space of test functions
${\cal J}(\hat{x}^2)$ and should be understood as a limit of functions
${\cal F}(\hat{x}_1,\hat{x}_2)\in {\cal J}(\hat{x}^2)$,
${\cal F}(\hat{x}_1,\hat{x}_2)\rightarrow \tilde{{\cal F}}
(\hat{x}_1,\hat{x}_2)$. The ideal EPR pair
correspond to the generalized state vector
$|\psi\rangle_{EPR}\in \Omega^*$ of the form
\begin{equation}
|\psi\rangle_{EPR}= \frac{1}{(2\pi)^3}
\int_{V_m^+}
\frac{d{\bf k} }{2k^0}
\mbox{e}^{-2ik^0x^0}
a^+({\bf k})a^+({\bf -k})
|0\rangle.
\end{equation}
This state is an analogue of the ideal EPR state [1] in the non-relativistic
case for the composite system consisting of two particles 1 and 2,
\begin{equation}
|\psi\rangle_{EPR}= \frac{1}{(2\pi)^3}
\int d{\bf k}
|{\bf k}\rangle_1 \otimes |{\bf -k}\rangle_2 ,
\end{equation}
where $|{\bf k}\rangle_{1,2}$ are the generalized eigenvectors of the
momentum operator, and $|{\bf k}\rangle_{1,2}\in {\cal J}^*({\bf k})$
(${\cal J}^*({\bf k})$ is the distribution space conjugate to
${\cal J}({\bf k})$). Accordingly, in the position representation
the state is written as
\begin{equation}
|\psi\rangle_{EPR}= \frac{1}{(2\pi)^3}
\int d{\bf x}
|{\bf x}\rangle_1 \otimes |{\bf x}\rangle_2 ,
\end{equation}
where $|{\bf x}\rangle_{1,2}$ are the generalized eigenvectors of the position
operator, and $|{\bf x}\rangle_{1,2}\in {\cal J}^*({\bf x})$
(${\cal J}^*({\bf x})$. The Fourier transform is known to map the space of
distributions ${\cal J}^*({\bf k})$ onto ${\cal J}^*({\bf x})$.

Qualitatively, at the intuitive level, the state (9) with
$\tilde{{\cal F}}$ from (8) can be interpreted as the creation
of two particles with ${\bf x}_1={\bf x}_2$ at time $x^0$ from vacuum,
simultaneously at the entire space (because of the presence of a
factor $const({\bf x}_1+{\bf x}_2)$). The EPR state is essentially non-local.

The one-particle packet state of the quantum field can be written as
\begin{equation}
|\psi\rangle_{s}= \varphi^+(f) |0\rangle=
\int d\hat{x} f(\hat{x})\varphi^+(\hat{x})  |0\rangle =
\frac{1}{(2\pi)^{3/2}}
\int_{V_m^+} \frac{ d{}{{\bf k}} }{2k^0}
f({\bf k}) a^+({\bf k}) |0\rangle,
\end{equation}
where $f({\bf k})$ is the packet amplitude in the ${\bf k}$-representation.
The state is defined by the equivalence class to which the
function $f(\hat{x})$ belongs. Different functions $f(\hat{x})$ which have
the same values on the mass shell define the same states.

The non-relativistic analogue of the packet is the state
\begin{equation}
|\psi\rangle_{s}=
\int d{\bf k} f({\bf k})  |{\bf k}\rangle_s,
\end{equation}
belonging in the non-relativistic case to the Hilbert state space
of the particle to be teleported, $|\psi\rangle_{s}\in {\cal H}_s$.

Because of the existence of a common vacuum state in the quantum field theory,
the vector corresponding to the system ``EPR pair + teleported state'' should
be written as
\begin{equation}
|\Psi\rangle=\varphi^+(f) {\cal P}(\varphi^+,{\cal F}) |0\rangle=
\int\int\int d\hat{x}_1 d\hat{x}_2 d\hat{x}
{\cal F}(\hat{x}_1,\hat{x}_2) f(\hat{x})
\varphi^+(\hat{x}_1) \varphi^+(\hat{x}_2) \varphi^+(\hat{x})
|0\rangle,
\end{equation}

The existence of a common vacuum state in the relativistic quantum field
theory results in a fundamental difference between the relativistic and
non-relativistic cases. In contrast to the non-relativistic case where
$|\psi\rangle_s\otimes |\psi\rangle_{EPR}\in {\cal H}_s\otimes {\cal H}_{12}=
{\cal H}_{s}\otimes{\cal H}_{1}\otimes {\cal H}_{2}$, the three-particle
states of the quantum field $|\Psi\rangle\in \Omega\subset {\cal H}$.
(Of course, a different representation of the state space
${\cal H}=\oplus_n\mbox{Sym}\otimes^n{\cal H}_1$ as a direct sum of the
symmetrized tensor products of the one-particle Fock spaces introduces
no changes because of the existence of a common cyclic vacuum vector.)

In addition, in the quantum field theory the states are all essentially
non-localizable in the sense that, as it was already established long ago
(see e.g. Ref.[7]), it is impossible to construct a state with a compact
support in the ${\bf x}$-representation using the normalized functions
$f({\bf k})$ defined on the mass shell (although the states with
the fall off arbitrarily close to the exponential one at the infinity
can be constructed) [8]. In some cases one can perhaps approximately assume
that the states of a composite system localized to within
the exponential tails in distant spatial domains can be regarded
as the states defined in the tensor product of the corresponding state
spaces which formally have different vacuum states. However, this assumption
is certainly wrong if the composite systems in entangled states and
their measurements are to be considered. This is exactly the case in the
teleportation problem. Moreover, if the state space of a composite system is
described as a tensor product of the constituent system state spaces,
the microcausality principle (commutation relations) is inevitably
violated since the operators acting in different Hilbert spaces
(factors in the tensor product) are certainly always commuting
independently of their position in the Minkowski space; to be more precise,
the operators even do not ``know'' about each other.

Let us now construct the appropriate measurement. Since in the relativistic
case the states are also described by the rays in the Hilbert space
(just as in the non-relativistic quantum mechanics), the measurements are also
described by the positive operator valued identity resolutions.

In the non-relativistic case the measurement used in the teleportation
procedure is described by an identity resolution in
${\cal H}={\cal H}_{s}\otimes{\cal H}_{1}\otimes {\cal H}_{2}$.  Let
${\Theta}$ be a measurable space of possible outcomes with the measure
$d\theta$; then
\begin{equation}
I_{s12}=I_{s}\otimes I_{1} \otimes I_{2}=
\int_{\Theta} {\cal M}_{\cal H}(d\theta)=
I_{1} \otimes \int_{\Theta} {\cal M}_{2s}(d\theta)=
I_{1}\otimes I_{2s}.
\end{equation}
The measurement (15) is only performed on one of the particles in the EPR-pair
and the particle in the unknown state to be teleported while the second
particle in the EPR-pair (the factor $I_1$) is not involved in the
measurement itself. It is important for the teleportation procedure that
the identity resolution in the entire state space of the three subsystems
can be expressed as a tensor product of the corresponding identity resolutions
in ${\cal H}_1$ and ${\cal H}_{s2}={\cal H}_s\otimes{\cal H}_2$, which
implicitly assumes the access to the individual subsystems.

For the relativistic quantum field, the identity resolution in the
three-particle states subspace cannot be in any way represented as a tensor
product of the appropriate identity resolutions in the one-particle and
two-particle subspaces. Such a measurement should only be constructed
as a general identity resolution in the entire three-particle states subspace.

It is first instructive to examine the measurement used in the
teleportation of a one-particle packet in the non-relativistic case:
\begin{equation}
{\cal M}_{\cal H}(d\theta)=I_{1}\otimes {\cal M}_{2s}(d\theta)=
I_{1}\otimes
|\Phi_{\bf XP}\rangle_{2s} \mbox{ }_{2s}\langle\Phi_{\bf XP}|
\frac{d{\bf X} d{\bf P} }{(2\pi)^3},
\end{equation}
where the space of possible outcomes is the set
${\Theta}=\{{\bf X}\times{\bf P}\in {\bf R}_{\bf X}\times {\bf R}_{\bf P}\}$.

Here ${\bf X}$ is the sum of the particle positions
${\bf X}={\bf x}_2+{\bf x}_s$,
${\bf P}={\bf p}_2-{\bf p}_s$ is the difference of their momenta, and
\begin{equation}
|\Phi_{\bf XP}\rangle_{2s}=
\int d{}{\bf k} \mbox{e}^{i {\bf kX} }
|{}{\bf k}\rangle_{2}\otimes |{}{\bf k+P}\rangle_{s}.
\end{equation}
It is easy to check that ${\cal M}(d{\bf X} d{\bf P})$ is actually an identity
resolution in ${\cal H}_2\otimes {\cal H}_s$; indeed,
\begin{equation}
I_{2s}= I_{2}\otimes  I_{s}=
\int
|\Phi_{\bf XP}\rangle_{2s} \mbox{ }_{2s}\langle\Phi_{\bf XP}|
\frac{d{\bf X} d{\bf P} }{(2\pi)^3}=
\end{equation}
\begin{displaymath}
\int\int d{\bf k}_1 d{\bf k}_2
\left(
|{\bf k}_1\rangle_2 \otimes |{\bf k}_2\rangle_s
\right)
\left(
{}_s\langle{\bf k}_2| \otimes {}_2\langle{\bf k}_1|
\right).
\end{displaymath}

A similar identity resolution for the relativistic quantum field in the
subspace of two-particle states is
\begin{equation}
I= \int_{V_m^+} \int_{V_m^+}
\frac{d {\bf k}_1}{2k_1^0} \frac{d {\bf k}_2}{2k_2^0}
\left(
a^+({\bf k}_1)  a^+({\bf k}_2) |0\rangle
\right)
\left(
\langle 0| a^-({\bf k}_2) a^-({\bf k}_1)
\right).
\end{equation}
Let us first write down the analogue of the measurement (17) and then complete
it to the identity resolution in the subspace of three-particle states.
The corresponding measurement can be represented in the form
\begin{equation}
{\cal M}(d\theta)=
\end{equation}
\begin{displaymath}
\left(
\int\int
d\hat{\xi}_1 d\hat{\xi}_2
\Phi(\theta,\hat{\xi}_1,\hat{\xi}_2)
\varphi^+(\hat{\xi}_1) \varphi^+(\hat{\xi}_2) |0\rangle
\right)
\left(
\int\int
d\hat{\xi}'_1 d\hat{\xi}'_2
\Phi^*(\theta,\hat{\xi}'_1,\hat{\xi}'_2)
\langle 0| \varphi^-(\hat{\xi}'_1) \varphi^-(\hat{\xi}'_2)
\right)
d\theta,
\end{displaymath}
which should give the identity resolution (19), i.e.
\begin{equation}
I=\int {\cal M}(d\theta)=
\left(
\int\int
d\hat{\xi}_1 d\hat{\xi}_2
\varphi^+(\hat{\xi}_1) \varphi^+(\hat{\xi}_2) |0\rangle
\right)
\left(
\int\int
d\hat{\xi}'_1 d\hat{\xi}'_2
\langle 0| \varphi^-(\hat{\xi}'_1) \varphi^-(\hat{\xi}'_2)
\right)
\end{equation}
\begin{displaymath}
\left(
\int d\theta
\Phi(\theta,\hat{\xi}_1,\hat{\xi}_2)
\Phi^*(\theta,\hat{\xi}'_1,\hat{\xi}'_2)
\right),
\end{displaymath}
which implies the conditions
\begin{equation}
\Phi(\theta,\hat{\xi}_1,\hat{\xi}_2)=
\delta(\xi^0_1-\xi^0) \delta(\xi^0_2-\xi^0)
\Phi(\theta,\mbox{\boldmath $\xi$}_1,\mbox{\boldmath $\xi$}_2),
\end{equation}
\begin{displaymath}
\int d\theta
\Phi(\theta,\mbox{\boldmath $\xi$}_1, \mbox{\boldmath $\xi$}_2),
\Phi^*(\theta,\mbox{\boldmath $\xi$}'_1, \mbox{\boldmath $\xi$}'_2)=
\delta(\mbox{\boldmath $\xi$}_1-\mbox{\boldmath $\xi$}_2)
\delta(\mbox{\boldmath $\xi$}'_1-\mbox{\boldmath $\xi$}'_2).
\end{displaymath}
It should be emphasized that the time $\xi^0$ is the same for
$\hat{\xi}_1,\hat{\xi}_2$ and $\hat{\xi}'_1,\hat{\xi}'_2$. We shall not
dwell on the interpretation of the measurement (20) and not only that
this measurement can be considered as a non-local measurement in the position
representation performed at time $\xi^0$.

The conditions (19--22) are satisfied if $\Phi$ is chosen in the from
\begin{equation}
\Phi(\theta,\mbox{\boldmath $\xi$}_1, \mbox{\boldmath $\xi$}_2)=
\delta(\mbox{\boldmath $\xi$}_1-\mbox{\boldmath $\xi$}_2)
\mbox{e}^{i {\bf P}\mbox{\boldmath $\xi$}_1 },
\quad
\theta=({\bf X},{\bf P} ),
\end{equation}
where ${\bf X},{\bf P}$ have the same meaning as in the non-relativistic case.

Finally, one obtains
\begin{equation}
{\cal M}( d{\bf X}d{\bf P} )=
\end{equation}
\begin{displaymath}
\left(
\int_{V^+_m}
\frac{d {\bf k} }{  \sqrt{2k^{0}({\bf k})} \sqrt{2k^{0}({\bf k+P})} }
\mbox{e}^{i {\bf kP}-i( k^0({\bf k}) + k^0({\bf k+P}) )\xi^0 }
a^+({\bf k}) a^+({\bf k+P}) |0\rangle
\right)
\end{displaymath}
\begin{displaymath}
\left(
\int_{V^+_m}
\frac{d {\bf k}' }{ \sqrt{ 2k^0({\bf k'}) } \sqrt{ 2k^0({\bf k'+P}) } }
\mbox{e}^{i {\bf k'P}+i( k^0({\bf k'}) + k^0({\bf k'+P}) ) \xi^0 }
\langle 0|  a^-({\bf k'+P)  a^-({\bf k'}})
\right)
\frac{d{\bf X} d{\bf P} }{(2\pi)^3}.
\end{displaymath}

We shall further need the following identity resolution in the
subspace of one-particle states of the quantum field:
\begin{equation}
I_1=
\int_ {V^+_m}  \frac{d {\bf k} }{2k^0}
\left( a^+({\bf k}) |0\rangle \right)
\left( \langle 0| a^-({\bf k})  \right)=
\int d{\bf x}
\left( \varphi^+(\hat{x}) |0\rangle \right)
\left( \langle 0| \varphi^-(\hat{x})  \right).
\end{equation}

The complete measurement in the subspace of the states of the composite
system consisting of the EPR-pair and the packet to be teleported is
\begin{equation}
{\cal M}_{\cal H}(d\theta)=
\end{equation}
\begin{displaymath}
\int d{\bf x}
\left(
\int\int
d\mbox{\boldmath $\xi$}_1 d\mbox{\boldmath $\xi$}_2
\Phi( \theta, \mbox{\boldmath $\xi$}_1,\mbox{\boldmath $\xi$}_2 )
\varphi^+(\hat{\xi}_1) \varphi^+(\hat{\xi}_2) \varphi^+(\hat{x}) |0\rangle
\right)
\end{displaymath}
\begin{displaymath}
\left(
\int\int
d\mbox{\boldmath $\xi$}'_1 d\mbox{\boldmath $\xi$}'_2
\Phi^*( \theta, \mbox{\boldmath $\xi$}'_1,\mbox{\boldmath $\xi$}'_2 )
\langle 0|\varphi^-(\hat{x}) \varphi^-(\hat{\xi}'_1) \varphi^-(\hat{\xi}'_2)
\right)
d\theta=
\end{displaymath}
\begin{displaymath}
\int d{\bf x}
\left(
\int d\mbox{\boldmath $\xi$}
\mbox{e}^{-i {\bf P}\mbox{\boldmath $\xi$} }
\varphi^+(\hat{\xi}) \varphi^+(\hat{\xi}-{\bf X}) \varphi^+(\hat{x})
|0\rangle
\right)
\left(
\int d\mbox{\boldmath $\xi$}'
\mbox{e}^{i {\bf P}\mbox{\boldmath $\xi$}' }
\langle 0|
\varphi^-(\hat{\xi}') \varphi^-(\hat{\xi}'-{\bf X}) \varphi^-(\hat{x})
\right)
\frac{ d{\bf X} d{\bf P} }{(2\pi)^3}=
\end{displaymath}
\begin{displaymath}
\int d{\bf x}
\left(
\varphi^+(\hat{x}) |\Phi_{\bf XP}\rangle
\right)
\left(
\langle\Phi_{\bf XP}| \varphi^-(\hat{x})
\right)
\frac{ d{\bf X} d{\bf P} }{(2\pi)^3},
\end{displaymath}
where
\begin{equation}
|\Phi_{\bf XP}\rangle=
\int d\mbox{\boldmath $\xi$} \mbox{e}^{-i {\bf P}\mbox{\boldmath $\xi$} }
\varphi^+(\hat{\xi}) \varphi^+(\hat{\xi}-{\bf X})  |0\rangle.
\end{equation}
Remember that in the variables $\hat{\xi}_1,\hat{\xi}_2$,
$\hat{\xi}'_1,\hat{\xi}'_2$, and $\hat{\xi},\hat{\xi}'$ the quantity
$\xi^0$ has the same value. For symmetry, we retain the four-dimensional
notation for the variables.

It should be noted that because of the existence of the common cyclic
(vacuum) vector the identity resolution in the subspace of three-particle
states cannot be in any way represented as a tensor product of the
corresponding identity resolutions in the subspaces of one- and two-particle
states. Unlike the non-relativistic case, the identity resolution (26)
for the relativistic quantum field cannot be reduced to the form defined
by Eq.(16).

The measurement (26) corresponds to the situation where the observation is
only performed on the two particles of three, while the third particle is not
involved in the measurement.

The probability of obtaining an outcome in the neighbourhood
$d{\bf X}d{\bf P}$ of the point ${\bf XP}$ of the space of
possible outcomes is given by the standard expression
\begin{equation}
\mbox{Pr}\{ d{\bf X} d{\bf P} \}=
\mbox{Tr}\{  |\Psi\rangle\langle\Psi | {\cal M}_{\cal H}( d{\bf X} d{\bf P} )\}=
\end{equation}
\begin{displaymath}
\left(
\int d{\bf x}
| {\cal A}( {\bf x},{\bf XP} ) |^2
\right)
d{\bf X} d{\bf P},
\end{displaymath}
where the total transition amplitude ${\cal A}( {\bf x},{\bf XP} )$
is defined as
\begin{equation}
{\cal A}( {\bf x},{\bf XP} )=
\int\int\int
d\hat{x}'d{\bf x}_1 d\mbox{\boldmath $\xi$}
\mbox{e}^{ i {\bf P}\mbox{\boldmath $\xi$} }
f(\hat{x}')
\end{equation}
\begin{displaymath}
\langle 0 |
\varphi^-(\hat{\xi}) \varphi^-(\hat{\xi}-{\bf X}) \varphi^-(\hat{x})
\varphi^+(\hat{x}_1) \varphi^+(\hat{x}_1) \varphi^+(\hat{x}')
| 0 \rangle,
\end{displaymath}
where two coordinates $\hat{x}_1,\hat{x}_1$ belong to the
EPR-pair, the variables $\hat{x}$ and $\hat{x}'$ correspond to the
packet with the shape $f(\hat{x}')$ being teleported, and, finally, the
coordinates $\hat{\xi},\hat{\xi}-{\bf X}$ refer to the measurement.

It should be noted that for the relativistic quantum field
there exists no analogy for the expression (2). The knowledge
of measurement (an operator valued measure ${\cal M}(d\theta)$)
itself is not sufficient to tell what states is the quantum system in
after the measurement which gave a particular outcome. To answer
this question one should know the instrument (superoperator) generating
the indicated operator valued measure. However, the superoperator
cannot be uniquely recovered from the given operator valued measure.
Fortunately, in the non-relativistic quantum mechanics it is sufficient
to know only the measurement itself to completely
describe the state of the teleported particle [5]. The latter is explained
by the possibility of the representation of the measurement itself
(an operator valued measure) as a tensor product of the appropriate
identity resolutions in the subspaces of the states of constituent subsystems.

For the relativistic quantum field it is impossible to represent the
measurement as a tensor product of the form (1). Therefore, asking
what is the state of the teleported system after the measurement for the
quantum field is physically meaningless and one can only speak of the
transition amplitude of the field as a whole from one state to another.

The vacuum average in Eq. (29) is only determined by the quantum field
properties. For a free field the vacuum average is decoupled into the
pairwise averages [5,6] so that only six contributions to the transition
amplitude ${\cal A}$ arise.

The quantity $| {\cal A}( {\bf x},{\bf XP} ) |^2$ can be interpreted as
the probability of detecting the ``teleported'' particle at a point
${\bf x}$ at time $x^0$ if the measurement gave an outcome in the
neighbourhood $({\bf X,P}; {\bf XP}+d{\bf X}d{\bf P})$. Similarly, the quantity
$ {\cal A}( {\bf x},{\bf XP} ) $ is the transition amplitude for the
packet from the state with the shape $f({\bf x}')$ at time $x^{0'}$ to
the point ${\bf x}$ at time $x^{0}$. One has the following expression for
the amplitude:
\begin{equation}
{\cal A}( {\bf x},{\bf XP} )=
2
\int\int\int
d\hat{x}'d{\bf x}_1 d\mbox{\boldmath $\xi$}
\mbox{e}^{ i {\bf P}\mbox{\boldmath $\xi$} }
f(\hat{x}')
\end{equation}
\begin{displaymath}
\left\{
{\cal D}_m^+( \hat{x}_1-\hat{x}' )
\left[
{\cal D}_m^+( \hat{x}-\hat{\xi} )
{\cal D}_m^+( \hat{x}_1-\hat{\xi}+{\bf X} )+
{\cal D}_m^+( \hat{x}_1-\hat{\xi} )
{\cal D}_m^+( \hat{x}-\hat{\xi}+{\bf X} )
\right]
\right.
+
\end{displaymath}
\begin{displaymath}
\left.
2{\cal D}_m^+( \hat{x}-\hat{x}' )
{\cal D}_m^+( \hat{x}_1-\hat{\xi} )
{\cal D}_m^+( \hat{x}_1-\hat{\xi}+{\bf X} )
\right\},
\end{displaymath}
where ${\cal D}_m^+( \hat{x} )$ is the commutator distribution for a
free field with mass $m$,
\begin{equation}
{\cal D}^{\pm}_{m}(\hat{x})=\pm\frac{1}{i(2\pi)^{3/2}}
\int \mbox{e}^{i\hat{p}\hat{x}}\theta(\pm p_0)\delta(\hat{p}^2-m^2)
d\hat{p}=
\end{equation}
\begin{displaymath}
\frac{1}{4\pi}\varepsilon({x_0})\delta(\hat{x}^2)\mp
\frac{im}{8\pi\sqrt{\hat{x}^2}}\theta(\hat{x}^2)
\left[N_1(m\sqrt{\hat{x}^2})\mp i\varepsilon(x_0)J_1(m\sqrt{\hat{x}^2})\right]
\pm\frac{im}{4\pi^2\sqrt{-\hat{x}^2}}\theta(-\hat{x}^2)
K_1(m\sqrt{-\hat{x}^2}),
\end{displaymath}
\begin{displaymath}
\varepsilon({x_0})\delta(\hat{x}^2)\equiv
\frac{\delta(x_0-|{\bf x}|) - \delta(x_0+|{\bf x}|)}{2|{\bf x}|}.
\end{displaymath}
To within the exponential tails, the commutator function is zero
beyond the light cone and has a singularity on its surface
$\lambda^2=(\hat{x}-\hat{x}')^2=0$; outside the light cone
the ${\cal D}^{\pm}(\lambda)$-function decay exponentially
at the Compton length as $|\lambda|^{-3/4}\exp{(-m\sqrt{|\lambda|})}$ [5,6].
At a fixed point $\hat{x}$ contributing to the integral are only points
$\hat{x}'$ lying within the light cone issued from the point $\hat{x}$,
which actually follows from the microcausality principle and impossibility
of the faster-than-light field propagation.

The amplitude (30) is actually a distribution which should be smeared with
a test function to obtain a final result. It should also be noted that
the product of any number of positive- or negative-frequency functions
${\cal D}^{\pm}(\hat{x})$ (unlike the product of causal functions)
is again correctly defined as a distribution from ${\cal J}^*(\hat{x})$
since in the momentum representation all these functions have their
supports located in the front part of the light cone.

Since we are only interested in the relative probabilities of different
processes, we shall directly employ the expression (30) for the amplitude.

Because of the common vacuum vector, it is impossible to arrange a measurement
in which there are no contributions to the transition amplitude from the
processes which are irrelevent to the teleportation. Formally, the fraction
of all these irrelevant processes is 1/2. This circumstance has
a fundamental nature and cannot be circumvented by any
geometrical tricks in the experiment.

The commutator function ${\cal D}^+_m(\hat{x}-\hat{y})$ describes the creation
of a particle at point $\hat{x}$, its propagation, and destruction at point
$\hat{y}$ (for $y^0>x^0$) [5,6]
\begin{equation}
\langle 0 | \varphi^-(\hat{y}) \varphi^+(\hat{x}) | 0 \rangle =
-i {\cal D}^+_m(\hat{x}-\hat{y}).
\end{equation}
Further, the Lorentz-invariant scalar product
\begin{equation}
(\varphi^-(f),\varphi^+(g))=
\langle 0 | \varphi^-(f) \varphi^+(g) | 0 \rangle =
\int\int d\hat{x} d\hat{y} f^*(\hat{y}) {\cal D}^+_m(\hat{x}-\hat{y})
g(\hat{x})=
\end{equation}
\begin{displaymath}
\int\int d\hat{x}
d\hat{y} f^*(\hat{y}) \langle 0 |
\varphi^-(\hat{y}) \varphi^+(\hat{x})
| 0 \rangle
g(\hat{x})
\end{displaymath}
is interpreted as the amplitude of the packet transition from the state
with the ``shape'' $g(\hat{x})$ to the state with the ``shape'' $f(\hat{y})$.
Since the test functions $g(\hat{x})$ and $f(\hat{y})$ determine the
state of the field through their values on the mass shell only, it is
convenient to rewrite the amplitude in the form
\begin{equation}
(\varphi^-(f),\varphi^+(g))=
-\frac{i}{ (2\pi)^{3/2} }
\int d\hat{p} \theta(p^0)\delta(\hat{p}^2-m^2)
f^*(\hat{p})g(\hat{p}),
\end{equation}
where $f(\hat{p})$ and $g(\hat{p})$ are the four-dimensional
Fourier transforms of the functions $f(\hat{x})$ and $g(\hat{x})$,
\begin{equation}
f(\hat{p})=
\frac{1}{(2\pi)^{3/2}}
\int d\hat{x} \mbox{e}^{-i\hat{p}\hat{x} }f(\hat{x}).
\end{equation}
Integration over the mass shell in Eq. (34) yields
\begin{equation}
(\varphi^-(f),\varphi^+(g))=
-\frac{i}{(2\pi)^{3/2}}
\int_{V^+_m} \frac{d{\bf p}}{2p_0}
f^*({\bf p}) \mbox{e}^{ip^0 y^0 }
g({\bf p}) \mbox{e}^{-ip^0 x^0 }=
\end{equation}
\begin{displaymath}
\int\int d{\bf x} d{\bf y}
f^*({\bf y}) {\cal D}^+_m(\hat{x}-\hat{y}) g({\bf x}),
\end{displaymath}
where the commutator function is defined as
\begin{equation}
{\cal D}^+_m(\hat{x}-\hat{y})=
-\frac{i}{(2\pi)^{3/2}}
\int_{V^+_m} \frac{d{\bf p}}{2p_0}
\mbox{e}^{i[ {\bf p} ( {\bf x - y} )-ip^0 (x^0- y^0) ] }
\quad
\hat{x}=(x^0,{\bf x}), \quad
\hat{y}=(y^0,{\bf y}),
\end{equation}
and
\begin{equation}
f({\bf x})=
\frac{1}{(2\pi)^{3/2}}
\int d{\bf p} \mbox{e}^{-i{\bf px} }f({\bf p}).
\end{equation}
The values of the test functions
$f({\bf p}) \mbox{e}^{ip^0 y^0 }$ and $g({\bf p}) \mbox{e}^{-ip^0 x^0 }$
on the mass shell uniquely determine the state and are interpreted as
the packet shape in the momentum representation. In the position
representation the quantities $f({\bf x})$ and $g({\bf y})$  are interpreted
as the spatial shape of the packet at times $x^0$ and $y^0$.
Note that the factors $\mbox{e}^{ip^0 y^0 }$ and $\mbox{e}^{-ip^0 x^0 }$
refer to the packet shape (one could simply write
$\tilde{f}({\bf p})=f({\bf p}) \mbox{e}^{ip^0 y^0 }$ and similarly for
$g({\bf p})$) and have nothing to do with the  dummy integration
variable in Eq. (33). This representation is chosen because in this form
the Lorentz-invariant scalar product (36) has the meaning of the transition
amplitude from the state which at time $x^0$ has the spatial shape
$g({\bf x})$ to the state with the spatial shape $f({\bf x})$
by the time $y^0$. To within the exponentially decreasing tails at
the Compton length outside the light cones, the contributions to this
amplitude are only given by the points lying inside the light cones
issued from each point $\hat{x}=(x^0,{\bf x})$
($|{\bf x-y}|^2-|x^0-y^0|^2<0$) where the function $g({\bf x})$
is different from zero.

Then in a similar way Eq.(30) can be rewritten in the form
where the integration is only performed over the spatial coordinates
\begin{equation}
{\cal A}( {\bf x},{\bf XP} )=
2
\int\int\int
d {\bf x}'d{\bf x}_1 d\mbox{\boldmath $\xi$}
\mbox{e}^{ i {\bf P}\mbox{\boldmath $\xi$} }
f({\bf x}')
\end{equation}
\begin{displaymath}
\left\{
{\cal D}_m^+( \hat{x}_1-\hat{x}' )
\left[
{\cal D}_m^+( \hat{x}-\hat{\xi} )
{\cal D}_m^+( \hat{x}_1-\hat{\xi}+{\bf X} )+
{\cal D}_m^+( \hat{x}_1-\hat{\xi} )
{\cal D}_m^+( \hat{x}-\hat{\xi}+{\bf X} )
\right]
\right.
+
\end{displaymath}
\begin{displaymath}
\left.
2{\cal D}_m^+( \hat{x}-\hat{x}' )
{\cal D}_m^+( \hat{x}_1-\hat{\xi} )
{\cal D}_m^+( \hat{x}_1-\hat{\xi}+{\bf X} )
\right\},
\end{displaymath}
where the quantity $f({\bf x})$ has the meaning of the spatial shape of the
unknown packet to be teleported at time $x^0$ ($\hat{x}=(x^0,{\bf x})$).
The quantity $x^0$ appears as a parameter in the arguments of the commutator
functions. The rest variables $x^0_1$, $x^{0'}$, $\xi^0$ also appear
in the arguments in $\hat{x}_1$,  $\hat{x}^{'}_{1}$, $\hat{\xi}^{'}$
as parameters. This form is best suitable for interpretation.

For example, the first term in Eq. (39) yields the amplitude of the process
associated with the creation of a non-local EPR-pair state (formally,
instantaneously in the entire space, as indicated by the integral
over ${\bf x}_1$) at time $x^0_1$, propagation of the packet in an
unknown state which at time $x^0$ has the shape $f({\bf x})$ and
subsequent joint measurement (also non-local, the integral over
$\mbox{\boldmath $\xi$}$) at time $\xi^0$ performed on the particle in
the unknown state and one of the particles of the EPR-pair.
In addition, one of the factors describes the free propagation
of the second particle in the EPR-pair, which is not involved in the
measurement, to the point $\hat{x}=(x^{0},{\bf x})$.
The second term in Eq. (39) has a similar interpretation.
The last two terms describe the processes irrelevant to the teleportation.
They describe the contributions to the amplitudes corresponding to
the processes where the measurement affects only the two particles of the
EPR-pair while the particle whose state is to be teleported propagates freely.

Although the transition to the non-relativistic theory cannot be performed
literately, it is still interesting to mention the
formal algorithm for this transition: One should omit all the terms in
the transition amplitude associated with the particle permutations and
replace the commutator distributions ${\cal D}^-_{m}(\hat{x})$ by
ordinary $\delta$-functions. Note that because of the singularity, this
replacement can only be understood symbolically.

The replacement of ${\cal D}_m^+( \hat{x} )$-functions
by ordinary $\delta({\bf x})$-functions is required because in the
non-relativistic case the integration is performed with the
Galilei-invariant measure $d\mu({\bf p})=d{\bf p}$, while in the
relativistic theory the Lorentz-invariant measure
$d\mu({\bf p})=\theta(p^0)\delta(\hat{p}^2-m^2)d\hat{p}       
=d{\bf p}/2p^0|_{V_m^+}$ is employed which finally gives
\begin{equation}
{\cal D}^+_m(\hat{x})= -\frac{1}{(2\pi)^{3/2}} \int_{V^+_m}
\frac{d{\bf p}}{2p_0} \mbox{e}^{i\hat{p}\hat{x}} \rightarrow
\frac{i}{(2\pi)^{3/2}}
\int d{\bf p}
\mbox{e}^{ i{\bf px} }=\delta({\bf x}).
\end{equation}
The temporal phase factors in the non-relativistic case do not matter
because of the absence of any limitations on the propagation speed.
Finally, the partial amplitude of the transition from point
${\bf x}'$ at time $x^{0'}$ to the point ${\bf x}$ at time $x^{0}$
we have
\begin{equation}
{\cal A}({\bf x}', {\bf x},{\bf XP} )=
\end{equation}
\begin{displaymath}
2 f({\bf x}) \mbox{e}^{ i {\bf P}{\bf x} }
\delta({\bf x}-{\bf x}'+{\bf X})
+
2 f({\bf x}) \mbox{e}^{ i {\bf P}({\bf x}+{\bf X}) }
\delta({\bf x}-{\bf x}'+{\bf X})+
4f({\bf x})\delta({\bf X})\delta({\bf P}),
\end{displaymath}
\begin{displaymath}
{\cal A}({\bf x},{\bf XP} )=
\int d{\bf x'}  {\cal A}({\bf x}', {\bf x},{\bf XP} ),
\end{displaymath}
where the partial amplitude for the transition
from point ${\bf x}'$ at time $x^{0'}$ to the point ${\bf x}$ at time $x^{0}$
``weighted'' with the packet shape $f({\bf x'})$ is introduced.

If the contribution of only the first term in Eq. (41) to
${\cal A}({\bf x}', {\bf x},{\bf XP} )$ is understood literally
as the amplitude of the transition to the point ${\bf x}$
under the condition that the measurement gave an outcome in the
interval $({\bf X,P}; {\bf XP}+d{\bf X}d{\bf P})$, this amplitude
coincides (to within an obvious unitary transformation which
is only determined by the measurement outcome, i.e. the value
of the pair ${\bf X,P}$) with the amplitude of propagation
of the wave packet having the shape $f({\bf x}')$ at the initial
moment of time $x^{0'}$ to the final point with coordinates
${\bf x}$,$x^{0}$
\begin{equation}
{\cal A}( {\bf x},{\bf XP} )=
f({\bf x}'-{\bf X}) \mbox{e}^{ i {\bf P}({\bf x}'-{\bf X}) }.
\end{equation}

For the probabilities (again understood symbolically) of obtaining different
measurement outcomes we have
\begin{equation}
\mbox{Pr}\{ d{\bf X} d{\bf P} \}=
\left(
\int d{\bf x}
| {\cal A}( {\bf x},{\bf XP} ) |^2
\right)
\frac{ d{\bf X} d{\bf P}}{(2\pi)^3}=
\end{equation}
\begin{displaymath}
\left(
\int d{\bf x}
| f( {\bf x} ) |^2
\right)
\frac{ d{\bf X} d{\bf P} }{ (2\pi)^3 }=
\frac{ d{\bf X} d{\bf P} }{ (2\pi)^3 };
\end{displaymath}
just as the ideal teleportation requires, the probabilities of obtaining
various measurement outcomes do not depend on the unknown state
which is to be teleported.

The second term in the amplitude (41) also refers to the teleportation
process where one of the particles of the EPR-pair and the particle in
the unknown state are exchanged (compared with the teleportation process
described by the first term in Eq. (41)).

The last term in Eq. (41) is irrelevant to the teleportation and arises
when the measurement only affects the two particles of the EPR-pair
(there are two equal contributions because of the exchange of the
particles within the EPR-pair). These processes contribute only
at the point ${\bf X}=0, {\bf P}=0$ of the outcome space
and their effect can in principle be eliminated by simply
discarding the measurements which gave this result.

Nevertheless, the first two terms in Eq. (41) describing the teleportation
process have different phase factors which does not allow to correctly
modify the transition amplitude by a unitary transformation similar to the
non-relativistic case (this would be possible if only the first term were
present).

At a first glance, one could simply keep only the measurements which gave
the results with ${\bf P}=0$ (when the phase factors are identical).
However, in that case the contribution of the ``parasitic'' processes
when the unknown packet is not affected by the measurement and propagates
freely becomes essential because of the $\delta$-functions
($\delta({\bf X})\delta({\bf P})$). Under these conditions it is
impossible to distinguish between the teleportation and free propagation
contributions to the transition amplitude.

In spite of the fact that the parasitic terms cannot be eliminated,
their contribution is only important in the vicinity of the point
${\bf X}=0, {\bf P}=0$ and has zero measure. Although being
rather strange at a first glance, this circumstance has a simple
qualitative interpretation related to the fact that each pure state
in the infinite-dimensional Hilbert space has in a certain sense
zero measure, as it is most simply explained in the non-relativistic
theory. The EPR-pair state is written as
\begin{equation}
|\psi\rangle_{EPR}=\frac{1}{(2\pi)^3}
\int d{\bf k} |{\bf k}\rangle_1 \otimes |{\bf k+q}\rangle_2,
\end{equation}
with a fixed ${\bf q}$. The parasitic terms correspond to the measurement
performed on the two particles of the EPR-pair which is actually reduced
to the projection on the state
\begin{equation}
|\Phi_{\bf XP}\rangle=\frac{1}{(2\pi)^3}
\int d{\bf k}^{'}
\mbox{e}^{i{\bf k^{'}X}}
|{\bf k}^{'}\rangle_1 \otimes |{\bf k^{'}+P}\rangle_2,
\end{equation}
which in fact is another EPR-pair with a different total momentum
(remember that we used the EPR-pair with ${\bf q}=0$). For the
projection we have
\begin{equation}
\langle \Phi_{\bf XP}|\psi \rangle_{EPR}
\propto
\delta({\bf P})
\int d{\bf k} \mbox{e}^{i{\bf kX}}=
\delta({\bf X})\delta({\bf P}),
\end{equation}
where the $\delta$-functions should be understood as indicated in the
above discussion.

The latter means that the measure of an individual EPR-pair with a fixed
${\bf q}$ among the entire set of all EPR-pairs is zero. Therefore,
the probability of occurrence of a particular EPR-pair with a specified
${\bf q}$ is zero (the measurement runs over the entire set of EPR-pairs).

Thus, in the relativistic quantum field theory the existence of a common
cyclic vector state together with the microcausality principle
(commutation relations) make the quantum teleportation impossible
in the sense it is understood in the non-relativistic quantum mechanics.

It should be emphasized once again that all the above arguments
are only applicable to the teleportation of a completely unknown
field state. In that case there is no way to ``label'' the individual
particles involved in the teleportation procedure. However, if the
state to be teleported is only partly unknown (e.g. for the case of
photon field only the polarization state is unknown while the photon
momentum and the total momentum of the EPR-pair are specified beforehand),
the available information can be used to construct the ``labels''
distinguishing the identical particles [9].

This work was supported by the Russian Foundation for Basic Research
(project No 99-02-18127), the project ``Physical Principles of the
Quantum Computer'',  and the program ``Advanced Devices and Technologies
in Micro- and Nanoelectronics''.

This work was also supported by the Wihuri Foundation, Finland.

\end{document}